\documentclass[12pt]{article}

\usepackage{mathtools}
\usepackage{amssymb,amsfonts}

\newcommand{\beq}{\begin{equation}}
\newcommand{\eeq}{\end{equation}}
\newcommand{\beqa}{\begin{eqnarray}}
\newcommand{\eeqa}{\end{eqnarray}}
\newcommand{\beqar}{\begin{eqnarray*}}
\newcommand{\eeqar}{\end{eqnarray*}}

\newcommand{\al}{\alpha}
\newcommand{\be}{\beta}

\def\spa          {\ \ \ }
\def\non          {\nonumber}
\def\ha           {\mbox{$\frac{1}{2}$}}

\def\spa          {\ \ \ }
\def\mand         {\spa\mbox{and}\spa}

\def\Tr           {\mbox{\rm Tr}\,}

\def\cd           {{\cdot}}
\def\ran          {\rangle}
\def\lan          {\langle}
\def\fsC    {C\!\!\!\!/\,}
\def\fsH    {H\!\!\!\!/\,}
\newcommand{\del}{\delta}

\newcommand{\eps}{\epsilon}
\newcommand{\ga}{\gamma}
\newcommand{\Ga}{\Gamma}

\newcommand{\inn}{\!\cdot\!}

\newcommand{\labell}[1]{\label{#1}} 
\newcommand{\reef}[1]{(\ref{#1})}

\def\sst#1{{\scriptscriptstyle #1}}
\def\0{{\sst{(0)}}}
\def\1{{\sst{(1)}}}
\def\2{{\sst{(2)}}}
\def\3{{\sst{(3)}}}
\def\4{{\sst{(4)}}}
\def\5{{\sst{(5)}}}
\def\6{{\sst{(6)}}}
\def\7{{\sst{(7)}}}
\def\8{{\sst{(8)}}}

\begin{document}
\baselineskip 18pt%
\begin{titlepage}
\vspace*{1mm}%
 

\vspace*{4mm}
\vspace*{4mm}%

\center{ {\bf \Large On BPS World Volume, RR Couplings \\
and their $\alpha'$ Corrections in  type IIB  
   }} 

\begin{center}
{Ehsan Hatefi }

\vspace*{0.4cm}{
Scuola Normale Superiore and INFN, Piazza dei Cavalieri, 7, 56126 Pisa, Italy \footnote{
ehsan.hatefi@sns.it, ehsanhatefi@gmail.com}\\
\quad\\

Faculty of Physics, University of Warsaw, ul. Pasteura 5, 02-093 Warsaw, Poland \\
\quad\\

Mathematical Institute, Faculty of Mathematics,
Charles University, P-18675, CR  }

\end{center}
\begin{center}{\bf Abstract}\end{center}
\begin{quote}

 We  compute the asymmetric and symmetric correlation functions of a four-point amplitude of a gauge field, a scalar field and a closed string Ramond-Ramond (RR) for different non-vanishing BPS branes. All world volume, Taylor and pull-back couplings and their all order  $\alpha'$ corrections have also been explored. Due to various symmetry structures, different restricted BPS Bianchi identities have also been constructed.  The prescription of exploring all the corrections of two closed string RR couplings in type IIB is given. We obtain  the closed form of the entire S-matrix elements of two closed string RR and a gauge field on the world volume of BPS branes in type IIB.  
 
 All the correlation functions of $< V_{A^{0}(x_1)}V_{C^{-1}(z_1,\bar{z}_1)}V_{C^{-1}(z_2,\bar{z}_2)}>$ are also revealed accordingly.  The algebraic forms for the most general case of the integrations $\int d^2z |z-i|^{a} |z+i|^{b} (z - \bar{z})^{c}
(z + \bar{z})^{d}$  
 on upper half plane are derived in terms of Pochhammer and some analytic functions. Lastly, we generate various singularity structures in both  effective field theory and IIB string theory, producing different contact interactions as well as their  $\alpha'$ higher derivative corrections.

 \end{quote}
\end{titlepage}

 \section{Introduction}
The fundamental objects are called $D_{p}$-branes that are long known to be existed. The late Joe Polchinski has given a life to the D-branes as dynamical objects and they are assumed to be sources for Ramond-Ramond (RR) closed string for all sorts of the stabilised BPS branes  \cite{Polchinski:1995mt,Witten:1995im}.

These RR couplings have various contributions in many different areas of theoretical high energy physics, ranging from pure String Theory to Mathematics, K-Theory as well as phenomenology. For example, one may  point out to the dissolving branes \cite{Douglas:1995bn} , K-theory
\cite{Minasian:1997mm,Witten:1998cd}, the well known Dielectric or Myers effect \cite{Myers:1999ps} where some of their $\alpha'$ corrections  are also derived in detail in \cite{Hatefi:2012zh}.

 To proceed with their dynamics, one needs to know all sorts of effective actions where various potentially interesting references  are given in \cite{Hatefi:2016wof}.  We would like to take advantage of  Conformal Field Theory (CFT) and try to release more information about the structures of the BPS effective actions. Indeed one of our aims is to work out with CFT to get more data and increase  our knowledge of deriving various string theory couplings and likewise various techniques to  Effective Field Theory (EFT) couplings along the way can also be explored.
 
 \vskip.1in
 
 One can just mention different applications to some of the known couplings, like the famous  $N^3$ phenomenon for $M5$ branes, dS solutions as well as entropy growth \cite{Hatefi:2012bp}. It is also known that RR plays the key role for all kinds of BPS and unstable branes   \cite{Hatefi:2010ik} where one can study some of its analysis as well as its dynamics in \cite{Kennedy:1999nn,Michel:2014lva}. All the standard approaches to get to EFT couplings were explained in detail in \cite{Hatefi:2012wj}, where S-matrix computations play the most fundamental role in getting the exact form of string couplings and their precise coefficients are even computed in the presence of higher derivative $\alpha'$ corrections.
We would like to illustrate just some of the BPS string calculations as given in \cite{Fotopoulos:2001pt}. For the sake of  comprehensiveness and for a review of open strings as well as their properties, we just highlight the original papers that are known and appeared in \cite{orientifold}.


\vskip.2in

In the first part of the paper, we calculate both asymmetric and symmetric S-matrices of a four point amplitude of a gauge, a scalar field and a closed string RR for different non-vanishing traces of all different BPS branes. All world volume, Taylor and pull-back couplings and their $\alpha'$ corrections have also been figured out. Due to various symmetry structures, various restricted BPS Bianchi identities can also be derived. We then try to demonstrate a prescription of exploring all the corrections of two closed string RR couplings in type IIB as well. To proceed further, we go head to obtain the closed form of the entire S-matrix elements of two closed string RR's and a gauge field on the world volume of BPS branes just in type IIB string theory. 

 \vskip.2in
 
 We first derive all the correlation functions of $< V_{A^{0}(x_1)}V_{C^{-1}(z_1,\bar{z}_1)}V_{C^{-1}(z_2,\bar{z}_2)}>$ and then reveal  the algebraic forms of the integrals for the most general integrations on the upper half plane that are of the sort of $\int d^2z |z-i|^{a} |z+i|^{b} (z - \bar{z})^{c}
(z + \bar{z})^{d}$  
 and  the outcome is written down in terms of Pochhammer and some analytic functions. Finally, we try to reconstruct  various singularity structures in both  Effective Field Theory (EFT) and IIB string theory.  We also work out with different contact term analysis and reproduce different contact interactions of the same S-matrix as well as their  $\alpha'$ higher derivative corrections and lastly point out to different remarks as well.

 \vskip.2in

The lower order supersymmetric generalisation of  Wess-Zumino (WZ) action has been found in \cite{Hatefi:2016fvm} in both IIB and IIA. It was also highlighted that all structures as well as the coefficients of the $\alpha'$ corrections in type IIB are different from their type IIA  couplings. 

\vskip.1in

It is worth taking into account the reference  \cite{Sen:2015hia} that deals with potentially different areas where important remarks on perturbative string amplitude calculations have been given. Indeed to do so, a systematic setup was revealed. It is highlighted that to ignore  some spurious singularities one needs to employ the vertical integration formalism with great care. Although this conjecture has potential overlaps with string calculations of IIB analysis, however, the role of the world volume couplings is not clarified in detail,  nor the bulk singularity structures are given. In the upcoming paper \cite{Hatefi:201899}  we address in detail the world volume and bulk singularity analysis of IIA.  In this paper, we just would like to show the method of deriving algebraic forms of all the integrals in terms of Pochhammer and consequently argue about the role of the world volume couplings just in IIB as well as their explicit corrections.

\vskip.1in

\section{All order  Corrections to $<V_{C^{-2}} V_{\phi^{0}}V_{A^{0}}>$ }

In this section using direct CFT methods \cite{Friedan:1985ge} we would like to explore the entire  S-matrix  elements of an asymmetric RR, a scalar field and a gauge field where its all order  $\alpha'$ higher derivative corrections can also be examined accordingly. It will be given by  exploring all its correlation function  as
\beqa
{\cal A}^{C^{-2} \phi^{0} A^{0}} & \sim & \int dx_1dx_2d^2z
 \lan
V_{\phi}^{(0)}(x_1)V_{A}^{(0)}(x_2)
V_{RR}^{(-2)}(z,\bar{z})\ran,\labell{cor11}\eeqa
The related vertex operators are read off  from \cite{Bianchi:1991eu}
  and \cite{Liu:2001qa} as follows
\beqa
V_\phi^{(-1)}(x)&=&e^{-\phi(x)}\xi_{1i}\psi^i(x)e^{ \alpha'iq\inn X(x)}\nonumber\\
V_A^{(-1)}(x)&=&e^{-\phi(x)}\xi_a\psi^a(x)e^{ \alpha'iq\inn X(x)}\nonumber\\
V_{\phi}^{(0)}(x) &=& \xi_{1i}(\partial^i X(x)+i\alpha'q.\psi\psi^i(x))e^{\alpha'iq.X(x)}\nonumber\\
V_{A}^{(0)}(x) &=& \xi_{1a}(\partial^a X(x)+i\alpha'q.\psi\psi^a(x))e^{\alpha'iq.X(x)} \nonumber\\
V_{C}^{(-\frac{3}{2},-\frac{1}{2})}(z,\bar{z})&=&(P_{-}\fsC_{(n-1)}M_p)^{\alpha\beta}e^{-3\phi(z)/2}
S_{\al}(z)e^{i\frac{\alpha'}{2}p\cd X(z)}e^{-\phi(\bar{z})/2} S_{\be}(\bar{z})
e^{i\frac{\alpha'}{2}p\cd D \cd X(\bar{z})}\nonumber\\
V_{C}^{(-\frac{1}{2},-\frac{1}{2})}(z,\bar{z})&=&(P_{-}\fsH_{(n)}M_p)^{\alpha\beta}e^{-\phi(z)/2}
S_{\al}(z)e^{i\frac{\alpha'}{2}p\cd X(z)}e^{-\phi(\bar{z})/2} S_{\be}(\bar{z})
e^{i\frac{\alpha'}{2}p\cd D \cd X(\bar{z})}\label{vop1}\eeqa
\vskip .1in

Note that, the total  background charge of the world-sheet with topology of a disk must be  -2, hence one needs to consider the symmetric and asymmetric picture of RR as illustrated in \reef{vop1}. For our notation we use $\mu,\nu=0,1,..,9$ where world volume indices run by $a,b,c=0,1,...,p$ and finally transverse indices are represented by $i,j=p+1,...,9$. Other notations for spinors and projector are given by the following formulae
\begin{displaymath}
P_{-} =\ha (1-\ga^{11}), \quad
\fsH_{(n)} = \frac{a
_n}{n!}H_{\mu_{1}\ldots\mu_{n}}\ga^{\mu_{1}}\ldots
\ga^{\mu_{n}}, (P_{-}\fsH_{(n)})^{\al\be} =
C^{\al\del}(P_{-}\fsH_{(n)})_{\del}{}^{\be}.
\non\end{displaymath}
where in type IIA  (type IIB) field strength of RR takes the value of $n=2,4$,$a_n=i$  ($n=1,3,5$,$a_n=1$). In order to use the holomorphic parts of the world-sheet fields, we apply the doubling trick which means that, some change of variables is taken into account
\begin{displaymath}
\tilde{X}^{\mu}(\bar{z}) \rightarrow D^{\mu}_{\nu}X^{\nu}(\bar{z}) \ ,
\spa
\tilde{\psi}^{\mu}(\bar{z}) \rightarrow
D^{\mu}_{\nu}\psi^{\nu}(\bar{z}) \ ,
\spa
\tilde{\phi}(\bar{z}) \rightarrow \phi(\bar{z})\,, \mand
\tilde{S}_{\al}(\bar{z}) \rightarrow M_{\al}{}^{\be}{S}_{\be}(\bar{z})
 \ ,
\non\end{displaymath}
and the following matrices are needed
\begin{displaymath}
D = \left( \begin{array}{cc}
-1_{9-p} & 0 \\
0 & 1_{p+1}
\end{array}
\right) \ ,\,\, \mand
M_p = \left\{\begin{array}{cc}\frac{\pm i}{(p+1)!}\ga^{i_{1}}\ga^{i_{2}}\ldots \ga^{i_{p+1}}
\eps_{i_{1}\ldots i_{p+1}}\,\,${ for $p$ even}$\\ \frac{\pm 1}{(p+1)!}\ga^{i_{1}}\ga^{i_{2}}\ldots \ga^{i_{p+1}}\ga_{11}
\eps_{i_{1}\ldots i_{p+1}} \,\,${for $p$ odd}$\end{array}\right.
\non\end{displaymath}
\vskip .2in
Now one is able to pick up just the following propagators for the whole world sheet fields of the kind of $X^{\mu},\psi^\mu, \phi$, as 
\begin{eqnarray}
\lan X^{\mu}(z)X^{\nu}(w)\ran & = & -\eta^{\mu\nu}\log(z-w) \ , \non \\
\lan \psi^{\mu}(z)\psi^{\nu}(w) \ran & = & -\eta^{\mu\nu}(z-w)^{-1} \ ,\non \\
\lan\phi(z)\phi(w)\ran & = & -\log(z-w) \ .
\labell{prop2}\end{eqnarray}

Replacing the related vertex operators inside \reef{cor11}, exploring correlation functions, fixing the SL(2,R) symmetry by gauge fixing as 
 $(x_1,x_2,z,\bar z)=(x,-x,i,-i)$ and  introducing $t = -\frac{\alpha'}{2}(k_1+k_2)^2$ 
one finds out the non zero part of the asymmetric amplitude as follows
\beqa
{\cal A}^{\phi^{0}A^{0}C^{-2}}&=& \bigg(-2ik_{2c}p^i\Tr(P_{-}\fsC_{(n-1)}M_p\Gamma^{ac})
 +2ik_{1b}k_{2c}\Tr(P_{-}\fsC_{(n-1)}M_p\Gamma^{acib})\bigg)\nonumber\\&&\times
 \mu_p \xi_{1i}\xi_{2a}\pi^{1/2}\frac{\Ga[-t+1/2]}{\Ga[1-t]}\label{lop33}\eeqa
\vskip.1in
where the traces can be explored, note that
\beqa
  p>3 , H_n=*H_{10-n} , n\geq 5.
  \nonumber\eeqa
and the compact form of the asymmetric amplitude is derived to be

\beqa
{\cal A}^{\phi^{0}A^{0}C^{-2}}&=& \bigg(-k_{2c}p^i\eps^{a_{0}\cdots a_{p-2}ac}C_{a_{0}\cdots a_{p-2}}
 +k_{1b}k_{2c}\eps^{a_{0}\cdots a_{p-3}acb}C^{i}_{a_{0}\cdots a_{p-3}}\bigg)\nonumber\\&&\times
 \mu_p \frac{32}{p!}\xi_{1i}\xi_{2a}\pi^{1/2}\frac{\Ga[-t+1/2]}{\Ga[1-t]}\label{lop33}\eeqa

We are dealing with massless strings, hence the expansion is low energy expansion\footnote{Note that we removed the over all factor $(2i)^{-2t-1}$}, that is,  $t=-p^ap_a \rightarrow 0$ and the expansion is found as follows

\beqa
\pi^{1/2}\frac{\Ga[-t+1/2]}{\Ga[1-t]}
 &=&\pi \sum_{m=-1}^{\infty}c_m t^{m+1},c_{-1}=1,c_0=2ln2,c_1=\frac{1}{6}\pi^2+2ln2
\ .\labell{taylor61}\nonumber
\eeqa

In order to produce the first term \reef{lop33} in an EFT, one needs to consider the mixed  Chern-Simons effective action and Taylor expansion of the scalar field as below
\beqa
S_1&=&\frac{\mu_p}{p!}(2\pi\alpha')^2 \int_{\Sigma_{p+1}} \partial_i C_{p-1}\wedge F\phi^i \labell{highaa}\eeqa
 
 Now if we consider  the covariant derivative of the scalar field from pull-back of brane and employ the following new effective action 
  \beqa
S_2&=&\frac{\mu_p}{p!}(2\pi\alpha')^2 \int_{\Sigma_{p+1}}C^{i}_{p-2}\wedge F\wedge D\phi_i \labell{highaa22}\eeqa
  then one can show that $S_2$ precisely produces the second term of \reef{lop33}.
 
 However, as can be observed the expansion of the amplitude consists of many contact interaction terms and one reconstructs all  the contact terms of the S-matrix in an EFT by imposing an infinite higher derivative corrections to the above $S_1$ and  $S_2$ effective actions. Therefore all contact terms for the first term of the asymmetric amplitude can be reconstructed by applying all order corrections to $S_1$ as follows:
 \beqa
\frac{\mu_p}{p!}(2\pi\alpha')^2 \int_{\Sigma_{p+1}}  \partial_i C_{p-1}\wedge\Tr \bigg( \sum_{n=-1}^{\infty}c_n(\alpha')^{n+1}
  D_{a_1}\cdots D_{a_{n+1}}FD^{a_1}...D^{a_{n+1}}\phi^i\bigg)
 \labell{highaa33}\eeqa

Likewise all order extensions of  $S_2$ are read off
 \beqa
\frac{\mu_p}{p!}(2\pi\alpha')^2 \int_{\Sigma_{p+1}}C^{i}_{p-2}\wedge
 \Tr(\sum_{n=-1}^{\infty}c_n(\alpha')^{n+1}  D_{a_1}\cdots D_{a_{n+1}}
 F\wedge   D^{a_1}...D^{a_{n+1}} D\phi_i)\nonumber\eeqa

\vskip.2in


On the other hand, the symmetric result of the amplitude  $<V_{C^{-1}} V_{\phi^{-1}}V_{A^{0}}>$ can be revealed as follows
\beqa
 {\cal A}^{\phi^{-1}A^{0}C^{-1}}&=&2^{1/2}i\xi_{1i}\xi_{2a} k_{2b}\int_{-\infty}^{\infty}dx (1+x^2)^{2t-1} (2x)^{-2t}\Tr(P_{-}\fsH_{(n)}M_p\Gamma^{abi})\label{mty777}\eeqa

 One can also read off  $<V_{C^{-1}} V_{\phi^{0}}V_{A^{-1}}>$  accordingly as follows
\beqa
2^{1/2}\xi_{1i}\xi_{2a}\int_{-\infty}^{\infty}dx (1+x^2)^{2t-1} (2x)^{-2t}
 \bigg(k_{1b}\Tr(P_{-}\fsH_{(n)}M_p\Gamma^{bai})-p^i\Tr(P_{-}\fsH_{(n)}M_p\gamma^{a})\bigg) \nonumber\eeqa

Using momentum conservation along the world volume of brane  $(k_1+k_2+p)^a=0$, due to symmetry structures and to get consistent result
for both above symmetric amplitudes, one gets to derive the following restricted Bianchi identity for RR's field strength as below
 \beqa
 p_b\eps^{a_{0}\cdots a_{p-2}ba} H^i_{a_{0}\cdots a_{p-2}}+p^i\eps^{a_{0}\cdots a_{p-1}a} H_{a_{0}\cdots a_{p-1}} &=&0
 \label{majid}\eeqa 

In the next section we would like to deal with more complicated analysis.

\vskip.2in

 \section{ S-matrix of $ \lan V_{A}(x_1)V_{C}(z_1,\bar{z}_1) V_{C}(z_2,\bar{z}_2)\ran$ of type IIB}
 
 The complete form of S-matrix element of two closed string RR field   $ \lan
V_{C}^{(-1)}(x_1,x_2) V_{C}^{(-1)}(x_4,x_5)\ran$ has been carried out in \cite{Hatefi:2016fvm} which has the following form
\beqa
&&\int dx_1 dx_2 dx_4 dx_5  (P_{-}\fsH_{(1n)}M_p)^{\alpha\beta}(P_{-}\fsH_{(2n)}M_p)^{\gamma\delta}(x_{12}x_{14}x_{15}x_{24}x_{25}x_{45})^{-1}
 \nonumber\\&&\times \frac{1}{2} \bigg[(\gamma^\mu C)_{\al\be}(\gamma^\mu C)_{\gamma\delta} x_{15} x_{24}-(\gamma^\mu C)_{\gamma\be}(\gamma^\mu C)_{\al\delta} x_{12} x_{45}\bigg]\nonumber\\&&\times
|x_{12}x_{45}|^{-s/2}|x_{14}x_{25}|^{-t/2}|x_{15}x_{24}|^{(s+t)/2}\nonumber\eeqa

It is shown that by choosing the gauge fixing as $(x_1,x_2,x_4,x_5)=(iy,-iy,i,-i)$ for the moduli space , one maps the moduli space to unit disk and the final amplitude of two closed string RR's in IIB was read off as follows

\beqa
{\cal A}^{CC}_{IIB} &=& \frac{i\mu_{1p}\mu_{2p}}{ p!p!}
\bigg(\frac{2}{s}-t\sum_{n,m=0}^{\infty} h_{n,m} (ts)^n(t+s)^m\bigg)\nonumber\\&&\times
\eps_{1}^{a_{0}\cdots a_{p-1}a}H_{1a_{0}\cdots a_{p-1}}\eps_{2}^{a_{0}\cdots a_{p-1}a}H_{2a_{0}\cdots a_{p-1}}
\label{esi99ccx}\eeqa

It was also shown that the only  massless pole can be reconstructed by using the following sub amplitude in an EFT
\beqa
{\cal A}&=&V^a_{\alpha}(C_{1p-1},A)G^{ab}_{\alpha\beta}(A)V^b_{\beta}(C_{2p-1},A)\labell{amp44390}
\eeqa

where the vertex $V^a_{\alpha}(C_{1p-1},A)$  is  derived from 
$i(2\pi\alpha')\mu_{1p}\int_{\Sigma_{p+1}}C_{1p-1}\wedge F$ and the following vertices were needed 

\beqa
V^a_{\alpha}(C_{1p-1},A)&=&i(2\pi\alpha')\frac{\mu_{1p}}{p!} \eps_{1}^{a_{0}\cdots a_{p-1}a}H_{1a_{0}\cdots a_{p-1}}
  \Tr( \lambda_\alpha)
\nonumber\\
G_{\alpha\beta}^{ab}(A)&=&\frac{-1}{(2\pi\alpha')^2}\frac{\delta^{ab}
\delta_{\alpha\beta}}{k^2}\,\,\, \nonumber\\
V^b_{\beta}(C_{2p-1},A)&=&i(2\pi\alpha')\frac{\mu_{2p}}{p!} \eps_{2}^{a_{0}\cdots a_{p-1}b}H_{2a_{0}\cdots a_{p-1}}
  \Tr( \lambda_\beta)
\label{ver138}
\eeqa
 $k^2=(p_1+D.p_1)^2=-s$  is taken inside the propagator while $t=-2p_1.p_2$.  Due to having the entire and closed form of S-matrix of two RR's in \reef{esi99ccx}
  one can apply properly higher derivative corrections on $C_1$ and $C_2$ so that  $(ts)^n$ can be produced by the following all order $\alpha'$ corrections to two closed string RR of  IIB 
\beqa
&&\sum_{n,m=0}^{\infty} h_{n,m} (\alpha')^{2n+1}  \bigg(D_{a_{1}}...D_{a_{n+1}} (D_bD_b)^n C_{1a_0\cdots a_{p-2}}  D^{a_{1}}...D^{a_{n+1}} C_{2a_0\cdots a_{p-2}}\bigg)\nonumber\\&&\times
\frac{\mu_{1p} \mu_{2p}}{(p-1)!(p-1)!}\eps_{1}^{a_{0}\cdots a_{p-2}a}\eps_{2}^{a_{0}\cdots a_{p-2}a}\label{5gh}
\eeqa

 $(t+s)^m$ can also be produced by the following all order $\alpha'$ corrections of  IIB 
\beqa
&&(\alpha'/2)^{m}  \bigg((D_aD^a) C_{1a_0\cdots a_{p-2}} C_{2a_0\cdots a_{p-2}}+\alpha' D^d C_{1a_0\cdots a_{p-2}} D_d C_{2a_0\cdots a_{p-2}}\bigg)^m\nonumber
\eeqa


\vskip.1in

Now in this section we would like to carry out direct CFT methods to derive the entire S-matrix elements of two closed string RR's and a gauge field on the world volume of BPS branes in IIB. The aim is to explore singularity structures as well as $\alpha'$ corrections and also to see if the above prescription holds or not. Hence, the five point function $ \lan
V_{A}^{(0)}(x_1)V_{C}^{(-1)}(z_1,\bar{z}_1) V_{C}^{(-1)}(z_2,\bar{z}_2)\ran$   in type IIB string theory ($z_1=x_2+ix_3$, $z_2=x_4+ix_5$; $x_{ij}=x_i-x_j$):
is given by the following correlation functions
\begin{align}
\label{eq:CCA} 
I_{CCA}&=\int_{\mathbb{R}}dx_1\int_{\mathcal{H}^+}dx_2dx_3\int_{\mathcal{H}^+}dx_4dx_5\left(P_{-}\fsH_{(1n)}M_p\right)^{\alpha\beta}\left(P_{-}\fsH_{(2n)}M_p\right)^{\gamma\delta} \notag\\ 
&\times \left(x_{23}x_{24}x_{25}x_{34}x_{35}x_{45}\right)^{-\frac{1}{4}}\mathcal{I}\xi_{1a}\Big(\langle :S_\alpha(x_2): :S_\beta(x_3)::S_\gamma(x_4)::S_\delta(x_5):\rangle \mathcal{J}^a \notag\\ 
&+2ik_{1c}\langle:S_\alpha(x_2): :S_\beta(x_3)::S_\gamma(x_4)::S_\delta(x_5)::\psi^c\psi^a(x_1):\rangle \Big),
\end{align}
where $\mathcal{I}$, 
and $\mathcal{J}^a$ take the form
\begin{align}\label{eq:I}
\mathcal{I}&=\lvert x_{12}\rvert^{2k_1\cdot p_1}\lvert x_{13}\rvert^{2k_1\cdot p_1}\lvert x_{14}\rvert^{2k_1\cdot p_2}\lvert x_{15}\rvert^{2k_1\cdot p_2}\lvert x_{23}\rvert^{p_1\cdot D\cdot p_1}\lvert x_{24}\rvert^{p_1\cdot p_2}\lvert x_{25}\rvert^{p_1\cdot D\cdot p_2}\notag\\ 
&\times \lvert x_{34}\rvert^{p_2\cdot D\cdot p_1}\lvert x_{35}\rvert^{p_1\cdot p_2}\lvert x_{45}\rvert^{p_2\cdot D\cdot p_2},
\notag\\ 
\mathcal{J}^a&=ip_1^a\frac{x_{52}}{x_{12}x_{15}}+ip_2^a\frac{x_{34}}{x_{14}x_{13}}.
\end{align}
with the definition of Mandelstam variables as
\begin{align}
s&=-(p_1+k_1)^2=-2k_1\cdot p_1 \\
v&=-(p_2+D\cdot p_2)^2=-2p_2\cdot D\cdot p_2 \\
w&=-(p_1+p_2)^2=-2p_1\cdot p_2.
\end{align}
Substituting the definition of Mandelstam variables into $\mathcal{I}$,  we obtain it as 
\begin{align}
\mathcal{I}=\lvert x_{12}\rvert^{-s}\lvert x_{13}\rvert^{-s}\lvert x_{14}\rvert^{s}\lvert x_{15}\rvert^{s}\lvert x_{23}\rvert^{2s-\frac{v}{2}}\lvert x_{24}\rvert^{-\frac{w}{2}}\lvert x_{25}\rvert^{-s+\frac{w+v}{2}}\lvert x_{34}\rvert^{-s+\frac{w+v}{2}}\lvert x_{35}\rvert^{-\frac{w}{2}}\lvert x_{45}\rvert^{-\frac{v}{2}}
\end{align}
The correlation function of four spin operators in IIB is read by
\begin{align}
&\mathcal{S}_{\alpha\beta\gamma\delta}(x_2,x_3,x_4,x_5):=\langle :S_\alpha(x_2): :S_\beta(x_3)::S_\gamma(x_4)::S_\delta(x_5):\rangle \notag\\
&=\bigg[(\gamma^\mu C)_{\alpha\beta}(\gamma_\mu C)_{\gamma\delta} x_{25}x_{34}
              -(\gamma^\mu C)_{\gamma\beta}(\ga_\mu C)_{\alpha\delta} x_{23} x_{45} \bigg]\frac{1}{2 (x_{23} x_{24} x_{25} x_{34} x_{35} x_{45})^{3/4}},
\end{align}
Correlation function of four spin operators and one current has been obtained in \cite{Hatefi:2013hca} to be 
\begin{align} 
&\widehat{\mathcal{S}}^{ca}_{\alpha\beta\gamma\delta}(x_1,x_2,x_3,x_4,x_5):=\langle:S_\alpha(x_2): :S_\beta(x_3)::S_\gamma(x_4)::S_\delta(x_5)::\psi^c\psi^a(x_1):\rangle \notag\\
&=\frac{(x_{23}  x_{24}  x_{25}  x_{34}  x_{35}  x_{45})^{-3/4}}{4 (x_{12}  x_{13}  x_{14}  x_{15})}\bigg[  \bigg((\gamma^c  C)_{\gamma \beta}  (\gamma^a C)_{\alpha \delta} -(\gamma^c  C)_{\alpha \delta}  (\gamma^a  C)_{\gamma \beta}\bigg)  (  x_{12}  x_{14}  x_{35}  - x_{15}  x_{13}  x_{24})
\notag\\
&\times x_{23}  x_{45}-\bigg( (\gamma^c  C)_{\alpha \beta}  (\gamma^a C)_{\gamma \delta} + (\gamma^c  C)_{\gamma \delta}  (\gamma^a  C)_{\alpha \beta}\bigg) x_{25}  x_{34} (  x_{15}  x_{13}  x_{24} + x_{14}  x_{12}  x_{35} ) \notag\\
& +   (\Gamma^{ca \lambda}  C)_{\alpha \beta}  (\gamma_\lambda  C)_{\gamma \delta}  x_{23}  x_{25}  x_{34} (  x_{14}  x_{15} )
+(\Gamma^{ca \lambda}  C)_{\gamma \delta}  (\gamma_\lambda  C)_{\alpha \beta}  x_{25} x_{34}  x_{45} (  x_{12}  x_{13} ) \notag\\
&-  (\Gamma^{ca \lambda}  C)_{\alpha \delta}  (\gamma_\lambda  C)_{\gamma \beta} x_{23}   x_{25}   x_{45} (  x_{14}   x_{13})
+  (\Gamma^{ca \lambda}  C)_{\gamma \beta}   (\gamma_\lambda  C)_{\alpha \delta}  x_{23}   x_{34}   x_{45}
 (  x_{12}  x_{15} )\bigg].           
\end{align}
Gauge fixing of $SL(2,\mathbb{R})$ invariance for $ \lan V_{A}^{(0)}(x_1)
V_{C}^{(-1)}(z_1,\bar z_1) V_{C}^{(-1)}(z_2,\bar z_2)\ran$   can be chosen as follows
\begin{align}
(z_1,\bar{z}_1,z_2,\bar{z}_2,x_1)=(i,-i,z,\bar{z},\infty).
\end{align}
The Jacobian for this transformation will be $Jac=-2ix_1^2$. After gauge fixing, the expressions for the amplitude in \eqref{eq:CCA} get simplified to
\begin{align}
{\cal A}^{A^{0}C^{-1}C^{-1}}&=(-2i) \int_{\mathcal{H}^+}dzd\bar{z} \left(P_{-}\fsH_{(1n)}M_p\right)^{\alpha\beta}\left(P_{-}\fsH_{(2n)}M_p\right)^{\gamma\delta} \big[(2i)\vert z+i\vert^2 \vert z-i\vert^2(z-\bar{z})\big]^{-\frac{1}{4}} \notag\\
&\times\mathcal{I}^{(g.f.)}\xi_{1a}\Bigg(\mathcal{S}^{(g.f.)}_{\alpha\beta\gamma\delta}(z,\bar{z})\mathcal{J}_{A}^{(g.f.),a}+2ik_{1c}\,\widehat{\mathcal{S}}^{(g.f.),ca}_{\alpha\beta\gamma\delta}(z,\bar{z})\Bigg).
\end{align}
The various gauge fixed quantities appearing in the above amplitudes are summarised \footnote{ The Momentum conservation along the brane is $(2k_1+p_1+D.p_1+p_2+D.p_2)^a=0$, also note that $(k_1+p_1)^2=p_{2a}p_{2a},(k_1+p_2)^2=p_{1a}p_{1a}$ and as we expected the expansion is low energy expansion, $s=-p_{2a}p_{2a},p_{1a}p_{1a},v,w\rightarrow 0$.}
in the following formulae ($z=x+iy;\;y\geq 0$)
\begin{align}
\mathcal{I}^{(g.f.)}&=(2i)^{2s-\frac{v}{2}}\vert z-i\vert^{-w}\vert z+i\vert^{-2s+w+v}(z-\bar z)^{-\frac{v}{2}},
\\
\mathcal{J}_A^{(g.f.),a}&=ip_1^a(\bar z-i)+ip_2^a(-i-z)\nonumber\end{align}
Now if we apply on-shell condition  for the gauge field $k_1.\xi_1=0$, then one gets to derive $\mathcal{J}_A^{(g.f.),a}=\frac{i}{2} (z+\bar z)(p_1-p_2)^a,$
\begin{align}
\mathcal{S}^{(g.f.)}_{\alpha\beta\gamma\delta}(z,\bar{z})&=\frac{1}{2\Big[(2i)(z-\bar{z})\vert z+i\vert^2\vert z-i\vert^2\Big]^{\frac{3}{4}}}\bigg[\vert z+i\vert^2(\gamma^\mu C)_{\alpha\beta}(\gamma_\mu C)_{\gamma\delta} 
              -2i(z-\bar{z})(\gamma^\mu C)_{\gamma\beta}(\ga_\mu C)_{\alpha\delta} \bigg],
\nonumber\\
\widehat{\mathcal{S}}^{(g.f.),ca}_{\alpha\beta\gamma\delta}(z,\bar{z})&=\frac{1}{4\Big[(2i)(z-\bar{z})\vert z+i\vert^2\vert z-i\vert^2\Big]^{\frac{3}{4}}}\Bigg\{2i(\bar z-z)(2i+(\bar z-z))\Big[(\gamma^c  C)_{\gamma \beta}  (\gamma^a C)_{\alpha \delta}\nonumber\\& -(\gamma^c  C)_{\alpha \delta}  (\gamma^a  C)_{\gamma \beta}\Big] +(z+\bar z)\vert z+i\vert^2\Big[(\gamma^c  C)_{\alpha \beta}  (\gamma^a C)_{\gamma \delta} + (\gamma^c  C)_{\gamma \delta}  (\gamma^a  C)_{\alpha \beta}\Big]\nonumber\\&+2i\vert z+i\vert^2(\Gamma^{ca \lambda}  C)_{\alpha \beta}  (\gamma_\lambda  C)_{\gamma \delta}
+(z-\bar z)\vert z+i\vert^2(\Gamma^{ca \lambda}  C)_{\gamma \delta}  (\gamma_\lambda  C)_{\alpha \beta}\notag \\&+2i(\bar z-i)(z-\bar z)(\Gamma^{ca \lambda}  C)_{\alpha \delta}  (\gamma_\lambda  C)_{\gamma \beta} -2i(z+i)(z-\bar z)(\Gamma^{ca \lambda}  C)_{\gamma \beta}  (\gamma_\lambda  C)_{\alpha \delta}\Bigg\}.\end{align}

Let us define the following integral 
\begin{align}
\mathbf{A}\left[a,b,c,d\right]=\int_{\mathcal{H}^+}dzd\bar{z}\vert z+i\vert^a\vert z-i\vert^b (z-\bar z)^c (z+\bar z)^d 
\end{align}
where $a,b.c$ are written down in terms of Mandelstam variables and $d=0,1$ for this amplitude, hence the final result for the 
the amplitude $ \lan V_{A}^{(0)}(x_1)
V_{C}^{(-1)}(z_1,\bar z_1) V_{C}^{(-1)}(z_2,\bar z_2)\ran$ in IIB can be expressed as
\begin{align}
 {\cal A}^{A^{0}C^{-1}C^{-1}}&=-(2i)^{2s-v/2} \left(P_{-}\fsH_{(1n)}M_p\right)^{\alpha\beta}\left(P_{-}\fsH_{(2n)}M_p\right)^{\gamma\delta}\Bigg\{\frac{i}{4}\xi_{1a}(p_1-p_2)^a(\gamma^\mu C)_{\alpha\beta}(\gamma_\mu C)_{\gamma\delta}\mathbf{A}_5 \notag\\
&+\frac{1}{2}\xi_{1a}(p_1-p_2)^a(\gamma^\mu C)_{\gamma\beta}(\gamma_\mu C)_{\alpha\delta}\mathbf{A}_6-k_{1c}\xi_{1a}\Bigg[ \Big[(\gamma^c  C)_{\gamma \beta}  (\gamma^a C)_{\alpha \delta} -(\gamma^c  C)_{\alpha \delta}  (\gamma^a  C)_{\gamma \beta}\Big] \notag\\
&\times\Big(-2i\mathbf{A}_2 +\mathbf{A}_4\Big)-\frac{i}{2}\Big[(\gamma^c  C)_{\alpha \beta}  (\gamma^a C)_{\gamma \delta} + (\gamma^c  C)_{\gamma \delta}  (\gamma^a  C)_{\alpha \beta}\Big]\mathbf{A}_5+(\Gamma^{ca \lambda}  C)_{\alpha \beta}  (\gamma_\lambda  C)_{\gamma \delta}\mathbf{A}_1 \notag\\
&-\frac{i}{2}(\Gamma^{ca \lambda}  C)_{\gamma \delta}  (\gamma_\lambda  C)_{\alpha \beta}\mathbf{A}_3+\frac{1}{2}(\Gamma^{ca \lambda}  C)_{\alpha \delta}  (\gamma_\lambda  C)_{\gamma \beta}\Big(\mathbf{A}_6-\mathbf{A}_4-2i\mathbf{A}_2\Big) \notag\\
&-\frac{1}{2}(\Gamma^{ca \lambda}  C)_{\gamma \beta}  (\gamma_\lambda  C)_{\alpha \delta}\Big(\mathbf{A}_6+\mathbf{A}_4+2i\mathbf{A}_2\Big)\Bigg]\Bigg\}.
\end{align} 
where ${A}_1,{A}_2,{A}_3,{A}_4,{A}_5,{A}_6$ are given by
\begin{align*}
\mathbf{A}_1&=\mathbf{A}\left[-2s+w+v,-w-2,-\frac{v}{2}-1,0\right] \\
\mathbf{A}_2&=\mathbf{A}\left[-2(s+1)+w+v,-w-2,-\frac{v}{2},0\right] \\
\mathbf{A}_3&=\mathbf{A}\left[-2s+w+v,-w-2,-\frac{v}{2},0\right] \\
\mathbf{A}_4&=\mathbf{A}\left[-2(s+1)+w+v,-w-2,-\frac{v}{2}+1,0\right] \\
\mathbf{A}_5&=\mathbf{A}\left[-2s+w+v,-w-2,-\frac{v}{2}-1,1\right] \\
\mathbf{A}_6&=\mathbf{A}\left[-2(s+1)+w+v,-w-2,-\frac{v}{2},1\right]. \\
\end{align*}
One can show that $A_5$ and $A_6$ have no contribution to our S-matrix , due to the fact that their integrations are zero on upper half plane. 

\section{ World-Volume Singularity Structures of IIB}

The  low energy expansions of all the functions can be found by using  the package of HypExp 
 \cite{Huber:2007dx} and we just point out to some of the expansions. For instance for $n=0$ ( see Appendix) and at first order of the expansion one gets the following values 
  \beqa
 \mathbf{A}_1&=& \frac{-i\pi (2s-v-w)}{2(2s-v)w}\\
 \mathbf{A}_2&=& \frac{-\pi (-2s+v)}{4(-2s+v+w)w}\\
 \mathbf{A}_3&=& \frac{\pi (-2s+v+w)}{(2s-v)w}\\
 \mathbf{A}_4&=& \frac{-i\pi (2s-v-2w)}{2(-2s+v+w)w}\\
 \nonumber\eeqa 
 
 Let us deal with the singularities of the S-matrix. The amplitude makes sense for $C_{p-3},C_{p-1},C_{p+1}$ cases. One can summarise the expansions of the functions in terms of  $A \epsilon series[n,\epsilon order]$ accordingly.
 \begin{table}[h]
\centering
\begin{tabular}{cc}
\hline
\quad\quad\quad$\epsilon$ order \quad\quad coefficient\quad \\
\hline
\quad\quad\quad-1\quad\quad\quad\quad 0\\
\\
\quad\quad\quad 0\quad\quad\quad\quad 0\\
\\
\quad\quad\quad\quad\quad\quad\quad\quad1\quad\quad  $\frac{1}{4} \pi^2(2s-v-w)$\\
\hline
\end{tabular}
\caption{A1$\epsilon series[1,1]$}
\end{table}
In particular, for the other cases we find the following expansions with the explicit coefficients as written in the tables 1-4.
 
 \begin{table}[h]
\centering
\begin{tabular}{cc}
\hline
\quad\quad\quad$\epsilon$ order \quad\quad coefficient\quad \\
\hline
\quad\quad\quad-1\quad\quad\quad\quad 0\\
\\
\quad\quad\quad 0\quad\quad\quad\quad 0\\
\\
\quad\quad1\quad\quad  $\frac{1}{8}i\pi^2 v$\\
\hline
\end{tabular}
\caption{A2$\epsilon series[1,1]$}
\end{table}
 
 \vskip.2in
 
 \begin{table}[h]
\centering
\begin{tabular}{cc}
\hline
\quad\quad\quad$\epsilon$ order \quad\quad coefficient\quad \\
\hline
\quad\quad\quad -1\quad\quad\quad\quad 0\\
\\
\quad\quad\quad 0\quad\quad\quad\quad 0\\
\\
\quad\quad\quad 1\quad\quad\quad\quad 0\\
\quad\quad\quad\quad\quad\quad\quad\quad\quad\quad\quad\quad 2\quad\quad\quad\quad $\frac{(i+1)}{4} \pi^2 v (-2s+v+w)$\\
\hline
\end{tabular}
\caption{A3$\epsilon series[1,2]$}
\end{table}

\vskip.1in

\begin{table}[h]
\centering
\begin{tabular}{cc}
\hline
\quad\quad\quad$\epsilon$ order \quad\quad coefficient\quad \\
\hline
\quad\quad\quad-1\quad\quad\quad\quad 0\\
\\
\quad\quad\quad 0\quad\quad\quad\quad $\frac{i\pi^2}{2}$\\
\\
\quad1\quad\quad  $\frac{-i\pi^2}{4} (v+EulerGamma  v +3i\pi v+ 4slog[2]+vPolyGamma[0,1/2])$\\
\hline
\end{tabular}
\caption{A4$\epsilon series[1,1]$}
\end{table}


\newpage

 One can show that the expansions of A3$\epsilon series[1,2]$, A3$\epsilon series[1,3]$ 
 and A3$\epsilon series[1,4]$ have no poles and made out of just contact interactions of the sort of the following form 

 \beqa 
(-2s+v+w) \Sigma_{m,p,q=0}^{\infty} h_{m,p,q} w^m v^ p s^q 
 \eeqa

Given the above structure, for $p=n$ case, now one can apply $\alpha'$ higher derivative corrections to explore  corrections in type IIB as follows
\beqa
&&\sum_{m,p,q=0}^{\infty} h_{m,p,q} (\alpha')^{m+q} \bigg(D^{a_{1}}...D^{a_{m}}  (D_aD^a)^p C_{2a_0\cdots a_{p-2}} D^{a_{1}}...D^{a_{q}} F_{a_{p-1}a_{p}} \nonumber\\&&\times 
D_{a_{1}}...D_{a_{m}} D_{a_{1}}...D_{a_{q}} C_{1a_0\cdots a_{p}} \bigg)
\frac{\mu_{1p} \mu_{2p}}{(p-1)!(p+1)!}\eps_{1}^{a_{0}\cdots a_{p}}\eps_{2}^{a_{0}\cdots a_{p}}\label{54gh}
\eeqa

where $(-2s+v+w)$ is an over all factor and $p_2.D.p_2$ can be constructed out by applying the sum of momenta  as 
$ \frac{1}{2} (D^aD_a) C_{2}\wedge F $. Note that for all the other functions starting from $n=3,\epsilon=0,1,2$ the only non zero values for the expansion would be at first order expansion and have the following non zero values  
  $\frac{\pi^2}{12} (2s-v-w)$,  $\frac{i\pi^2}{24}v$ and $\frac{i\pi^2}{12}v$
 for A1[3,1], A2[3,1] and A4[3,1] 
   accordingly.


  \vskip.2in

  Now given the low energy expansion and in order to produce the world volume singularity structures, we try to extract the traces and carry out algebraic simplifications.
  
Indeed if $\lambda$ in the S-matrix takes world volume index $\lambda=d$ then $p_1.D.p_1$
channel pole in IIB can produced by the following EFT sub amplitude 
 
 
  
\beqa
{\cal A}^{CCA}_{IIB} &=& \frac{i \pi \mu_{1p}\mu_{2_{p-2}}}{ p!(p-2)!}
 \frac{1}{2p_1.D.p_1}\eps_{1}^{a_{0}\cdots a_{p-1}d}H_{1a_{0}\cdots a_{p-1}}\eps_{2}^{a_{0}\cdots a_{p-3}cad}H_{2a_{0}\cdots a_{p-3}} \xi_{1a} k_{1c}
\label{esi99}\eeqa
  $ (\mu_{1p},\mu_{2_{p-2}})$ are RR charges and it is renormalised by $\frac{1}{2^6}$. Hence if  $\lambda$ picks up world volume index then all the traces have non-zero contributions for  $ C_{1_{p-1}}$ as well as for $C_{2_{p-3}}$ cases. Note that this obviously confirms that we do have a  gauge field singularity structure and also all various $\alpha'$ higher derivative corrections to two closed string RR in type IIB that can be constructed out later on.

\vskip.1in


The gauge singularity structure for this particular case is regenerated by the following EFT sub amplitude 
\beqa
{\cal A}&=&V^a_{\alpha}(C_{1p-1},A)G^{ab}_{\alpha\beta}(A)V^b_{\beta}(C_{2p-3},A,A_1),\labell{amp44390}
\eeqa
where  the Chern-Simons coupling 
$i(2\pi\alpha')\mu_{1p}\int_{\Sigma_{p+1}}C_{1p-1}\wedge F$ is needed, also note that it is shown that, this coupling does not receive any corrections either.
 
 
 \vskip.2in

One can reveal all the simple propagators by  employing the kinetic terms appeared in DBI action as $(2\pi\alpha')^2 F_{ab} F^{ab} $ and $\frac{(2\pi\alpha')^2}{2}  \Tr(D_a\phi^i D^a\phi_i)$ where the kinetic terms of gauge fields and scalars will receive no correction either, because they are fixed in the  low energy DBI action as well. One  readily gets to derive the EFT vertex operators for the above amplitude as follows  \beqa
V^a_{\alpha}(C_{1p-1},A)&=&i(2\pi\alpha')\frac{\mu_{1p}}{p!} \eps_{1}^{a_{0}\cdots a_{p-1}a}H_{1a_{0}\cdots a_{p-1}}
  \Tr( \lambda_\alpha)
\nonumber\\
G_{\alpha\beta}^{ab}(A)&=&\frac{-1}{(2\pi\alpha')^2}\frac{\delta^{ab}
\delta_{\alpha\beta}}{k^2}\,\,\, \nonumber\\
V^b_{\beta}(C_{2p-3},A,A_1)&=&i(2\pi\alpha')^2\frac{\mu_{2_{p-2}}}{(p-2)!} \eps_{2}^{a_{0}\cdots a_{p-1}b}H_{2a_{0}\cdots a_{p-3}} \xi_{1a_{p-2}} k_{1a_{p-1}}
  \Tr( \lambda_1\lambda_\beta)
\label{ver138yy}
\eeqa
where the following interaction for the 2nd Chern-Simons coupling has been taken into account
\beqa
i(2\pi\alpha')^2\mu_{2_{p-2}}\int_{\Sigma_{p+1}}C_{2_{p-3}}\wedge F \wedge F\eeqa
 Notice that in the propagator one considers $k^2=-(p_1+D.p_1)^2$. Now by replacing \reef{ver138yy} into \reef{amp44390}, 
we would be able to precisely reconstruct in the EFT of IIB the following simple pole 

\beqa
 \frac{i \pi \mu_{1p}\mu_{2_{p-2}}}{ p!(p-2)!}
 \frac{1}{2p_1.D.p_1}\eps_{1}^{a_{0}\cdots a_{p-1}d}H_{1a_{0}\cdots a_{p-1}}\eps_{2}^{a_{0}\cdots a_{p-1}d}H_{2a_{0}\cdots a_{p-3}} \xi_{1a_{p-2}} k_{1a_{p-1}}
\label{esi961nb}\eeqa

This is the same gauge field  singularity structure of  $CCA$ that appeared in IIB string theory.\footnote{
Notice that due to symmetries, likewise $p_2.D.p_2$ pole can also be generated. Given the fact that after all $\Tr(\lambda_1)$ is zero for SU(N) we come to know that there is no $s$ channel pole. $4k_1.p_2=-4k_1.p_1=2s$}
 It would be nice to explore two RR couplings with scalar field, their  bulk singularity structures to actually find out  their corrections in type IIA \cite{Hatefi:2018IIA}  as well.

  \vskip.2in

  One can show that indeed if $\lambda$ in the S-matrix takes the transverse index $\lambda=j$ then the scalar field pole in IIB is produced by the following EFT sub amplitude 
 
  
\beqa
  \frac{i \pi \mu_{1_{p+2}}\mu_{2_{p}}}{ (p+2)!(p)!}
 \frac{1}{2p_1.p_2}\eps_{1}^{a_{0}\cdots a_{p}}
 H^j_{1a_{0}\cdots a_{p}}
 \eps_{2}^{a_{0}\cdots a_{p-2}ca} H^j_{2a_{0}\cdots a_{p-2}} \xi_{1a} k_{1c}
\label{esi99rt}\eeqa
  Hence if  $\lambda$ picks up transverse index then,  the amplitude and the traces have non-zero contributions for  $ C_{1_{p+1}}$ as well as for $C_{2_{p-1}}$ cases and we do have a scalar field singularity structure that can be shown to be matched in an EFT.  

\vskip.1in


The singularity structure for this case is also produced by the following EFT  counterpart

\beqa
V^i_{\alpha}(C_{1p+1},\phi)G^{ij}_{\alpha\beta}(\phi)V^j_{\beta}(C_{2_{p-1}},\phi,A_1),\labell{amp4439055}
\eeqa

where $V^i_{\alpha}(C_{1_{p+1}},\phi)$  was obtained from $ i(2\pi\alpha')\mu_{1p}\int_{\Sigma_{p+1}} \partial^i C_{1_{p+1}}\phi_i $
 which is indeed the Taylor expansion in EFT. We clarify  the EFT vertex operators for the above amplitude as follows  
\beqa
V^i_{\alpha}(C_{1_{p+1}},\phi)&=&i(2\pi\alpha')\frac{\mu_{1_{p+2}}}{(p+2)!} \eps_{1}^{a_{0}\cdots a_{p}}H^{i}_{1a_{0}\cdots a_{p}}
  \Tr( \lambda_\alpha)
\nonumber\\
G^{ij}_{\alpha\beta}(\phi)&=&\frac{-1}{(2\pi\alpha')^2}\frac{\delta^{ij}
\delta_{\alpha\beta}}{k^2}\,\,\, \nonumber\\
V^j_{\beta}(C_{2_{p-1}},\phi,A_1)&=&i(2\pi\alpha')^2\frac{\mu_{2_{p}}}{(p)!} \eps_{2}^{a_{0}\cdots a_{p}}H^j_{2a_{0}\cdots a_{p-2}} \xi_{1a_{p-1}} k_{1a_{p}}
  \Tr( \lambda_1\lambda_\beta)
\label{ver138}
\eeqa
where the mixed WZ and CS interaction 
$i(2\pi\alpha')^2\mu_{2_{p}}\int_{\Sigma_{p+1}}\partial_i C_{2_{p-1}}\wedge F \phi^i$
 for the second coupling is taken.
Having set that \footnote{We have also used  the fact that $(k_1+p_2)^a=-p_1^a$ and also $2k_1.p_2=-2p_1.p_2$ in both effective field theory and string calculations.}, one regenerates $p_1.p_2$ singularity structure in EFT which is the same pole as appeared in \reef{esi99rt} of IIB.
Finally one can show that the other 
singularity of the amplitude can be reconstructed in an EFT.
 Indeed, if one makes use of the same EFT rule and applies the mixed pull back of brane and Chern-Simons coupling of  the derived form as below
  \beqa 
  \frac{i(2\pi\alpha')^2\mu_{1p}}{(p-1)!}\int_{\Sigma_{p+1}}C^i_{1_{p-2}}\wedge F \wedge D\phi^i \label{b21b}\eeqa
then one would be able to regenerate the singularity in an EFT side as well.

\section{Conclusion}

In this paper,  first we have computed both asymmetric and symmetric S-matrices of a four point amplitude of a gauge, a scalar field and a closed string RR for different non-vanishing traces of all different BPS branes.  We then figured out all world volume, Taylor and pull-back couplings as well as their $\alpha'$ corrections.  Thanks to symmetry structures, various restricted BPS Bianchi identities are also revealed. A prescription for the corrections of two closed string RR couplings in type IIB was found out. We have also gained the closed form of all the correlators of two closed string RR and a guage field in type IIB string theory.

 \vskip.2in
 
The algebraic forms of the integrals for the most general integrations on the upper half plane that are of the sort of $\int d^2z |z-i|^{a} |z+i|^{b} (z - \bar{z})^{c}
(z + \bar{z})^{d}$  are explored where the outcome is written down in terms of Pochhammer and some analytic functions. Lastly, various singularity structures in both EFT and IIB string theory are reconstructed.  We  have also worked out with different contact term analysis and reproduced different contact interactions of the S-matrices as well as their  $\alpha'$ higher derivative corrections and eventually some concrete points are clarified in detail.  Various world volume couplings just in IIB as well as their explicit corrections are discovered as well.

\vskip.1in

Now let us just address in detail  the technical issues related to solving integrals in the Appendix.
\vskip.2in

\section{Appendix 
\\
Solving the integrals of two RR's and an NS field}
 
We did gauge fixing as $(x_1 \rightarrow \infty, z_1=i,\bar z_1=-i, z_2=z,\bar z_2=\bar z)$ and eventually one needs to take integrations on 
the location of the second closed string on upper half plane as follows

\beqa
I = \int_{\cal{H}^{+}} d^2 \!z |z-i|^{a} |z+i|^{b} (z - \bar{z})^{c}
(z + \bar{z})^{d},
\nonumber\eeqa
where $d=0,1$ and $a,b,c$ are written down  in terms of three independent Mandelstam variables. We first use the following transformations 
\beqa
|z+i|^{b}= \frac{1}{\Gamma(- \frac{b}{2})} \int_{0}^{\infty} dt t^{-\frac{b}{2} - 1} e^{-t |i+z|^2}, \nonumber\\
|z-i|^{a}= \frac{1}{\Gamma(- \frac{a}{2})} \int_{0}^{\infty}
du u^{-\frac{a}{2} - 1} e^{-u |z-i|^2}. \nonumber
\nonumber\eeqa
where
 $z=x+iy$ and the integration on x is readily done as $\int^{\infty}_{-\infty} dx e^{-(t+u)x^2} =
\frac{\sqrt{\pi}}{(t+u)^{\frac{1}{2}}},$


Let us first solve it for $d=0$ so that the integration on $y$ becomes

\begin{eqnarray}
I_y= (2i)^c \int^{\infty}_{0} dy (y+\frac{u-t}{u+t})^c e^{-(t+u)(y)^2-\frac{4tu}{t+u}}.
\end{eqnarray}

Using simple algebraic analysis and change of variables one writes down the integration on y as follows
\begin{eqnarray}
I_y= (-2i)^c (\frac{t-u}{u+t})^{c+1}\int^{\infty}_{0} dy (1-y)^c e^{-\frac{(t-u)^2}{t+u}y^2} e^{-\frac{4tu}{t+u}}.
\end{eqnarray}

Now one can make use of Pochhammer definition as follows

\beqa
_1F_0({a}|z)=(1-z)^{-a}=\Sigma_{n=0}^{\infty} \frac{(a)_n z^n}{n!} ,  \int^{\infty}_{0} dy \; y^c e^{-(s+u)y^2} =
\frac{\Gamma(\frac{1+c}{2})}{2(s+u)^{\frac{1+c}{2}}}
\label{esi88}
\eeqa
If we consider \reef{esi88} then one gets to derive the whole y-integration. Now one can collect all the results of x and y integration inside $I$ to get to

\beqa
I=  \Sigma_{n=0}^{\infty} \frac{(-c)_n \Gamma(\frac{1+n}{2})}{n!}\frac{\sqrt{\pi}(-2i)^{c}} 
{2\Gamma(\frac{-a}{2}) 
\Gamma(\frac{-b}{2})}\int^{\infty}_{0}\int^{\infty}_{0}
\frac{dt du}{(t+u)^{c+1-n/2}} (t-u)^{c-n}u^{-a/2-1}t^{-b/2-1}e^{- \frac{4ut}{t+u}}\nonumber
\eeqa
If one uses the following change of variables
\beqa
u=\frac{x}{s},\quad t=\frac{x}{1-s},\quad dtdu=Jdxds=\frac{xdxds}{(s(1-s))^2}
\nonumber\eeqa

By substituting the above variables and making use of the Jacobian one eventually gets to gain the final integral as 
 \beqa
I&=& \int^{\infty}_{0}dxe^{-4x}x^{\frac{-4-(a+b+n)}{2}}\int^{1}_{0}ds  s^{(n+a)/2}(1-s)^{(n+b)/2}(1-2s)^{c-n}\nonumber\\&&\times
\Sigma_{n=0}^{\infty} \frac{(-c)_n (-1)^{c-n}\Gamma(\frac{1+n}{2})}{n!}\frac{\sqrt{\pi}(-2i)^{c}} 
{2\Gamma(\frac{-a}{2}) 
\Gamma(\frac{-b}{2})}\eeqa
 
If we use  $\Gamma(z)=(z-1)!$ then one reads off the final answer to be
 \beqa
I&=&\int^{1}_{0}ds  s^{(n+a)/2}(1-s)^{(n+b)/2}(1-2s)^{c-n}\nonumber\\&&\times
\Sigma_{n=0}^{\infty} \frac{(-c)_n (-1)^{2c-n}\Gamma(\frac{1+n}{2})}{n!}\frac{\sqrt{\pi}(2i)^{c}} 
{\Gamma(\frac{-a}{2}) 
\Gamma(\frac{-b}{2})} \Gamma( -1-
\frac{a+b+n}{2}) 2^{a+b+n+1}\eeqa

where the  integration $\int^{1}_{0}ds  s^{(n+a)/2}(1-s)^{(n+b)/2}(1-2s)^{c-n}$ can also be computed in terms of Hypergeometric function as follows

\beqa
&&\frac{(-2)^{c-n} \Gamma \left(\frac{1}{2} (b+n+2)\right) \Gamma \left(\frac{a}{2}+c-\frac{n}{2}+1\right) \, _2F_1\left(\frac{1}{2} (-a-b-2 c-2),n-c;\frac{1}{2} (-a-2 c+n);\frac{1}{2}\right)}{\Gamma \left(\frac{1}{2} (a+b+2 c+4)\right)}
\nonumber\\&&+\frac{ \left((-1)^c e^{\frac{1}{2} i \pi  (a+n+2)} \csc \left(\frac{1}{2} \pi  (a+2 c-n)\right)+(-1)^n \cot \left(\frac{1}{2} \pi  (a+n)\right)+i (-1)^n\right)}{\Gamma \left(-\frac{a}{2}-\frac{n}{2}\right) \Gamma \left(\frac{a}{2}+c-\frac{n}{2}+2\right)}
\nonumber\\&&\times
\pi  (-1)^{-n} 2^{-\frac{a}{2}-\frac{n}{2}-1} e^{-\frac{1}{2} i \pi  (a+n+2)} \Gamma (c-n+1) \, _2F_1\left(\frac{1}{2} (-b-n),\frac{1}{2} (a+n+2);\frac{1}{2} (a+2 c-n+4);\frac{1}{2}\right)\nonumber\eeqa

  \section*{Acknowledgments}

I would like to thank P. Vasko for discussions and for the expansions of the functions appeared in this paper. I am also grateful to A. Sagnotti, D. Francia, N. Arkani-Hamed, K. Narain, L. Alvarez-Gaume, P. Sulkowski, J. Balog, Z. Bajnok, J. Maldacena, A. Polyakov, Z. Bern, E. Witten and W. Siegel for very useful discussions. Some parts of the paper have been carried out at CERN, ICTP, Mathematical Institute at Charles university and Wigner Institute during my recent visits and I thank them for the great hospitality. This work is supported  by an ERC Starting Grant no. 335739
"Quantum fields and knot homologies", funded by the European Research Council under the European Union's 7th Framework Programme. I was also supported in part by Scuola Normale Superiore in Pisa and INFN (ISCSN4-GSS-PI), and by the MIUR-PRIN contract 2017CC72MK-003.

  \end{document}